\documentclass[manuscript]{aastex}

\shorttitle{test}
\shortauthors{Ohama et al. 2009}

\begin{document}
  

\title{Temperature and Density Distribution in the Molecular Gas Toward Westerlund 2: Further Evidence for Physical Association}


\author{A. Ohama,\altaffilmark{1} J. R. Dawson,\altaffilmark{1, 2} N. Furukawa,\altaffilmark{1}  A. Kawamura,\altaffilmark{1} N. Moribe,\altaffilmark{1} H. Yamamoto,\altaffilmark{1} T. Okuda,\altaffilmark{1} N. Mizuno,\altaffilmark{1,3} T. Onishi,\altaffilmark{1,4} H. Maezawa,\altaffilmark{5} T. Minamidani,\altaffilmark{6} A. Mizuno,\altaffilmark{5} $_{AND}$ Y. Fukui\altaffilmark{1}}
\affil{$^1$Department of Physics and Astrophysics, Nagoya University, Chikusa-ku, Nagoya, Aichi, 464-8601, Japan}
\affil{$^2$Australia Telescope National Facility, CSIRO, PO Box 67, Epping NSW 1710, Australia}
\affil{$^3$National Astronomical Observatory of Japan, Mitaka, Tokyo, 181-8588, Japan}
\affil{$^4$Department of Astrophysics, Graduate School of Science, Osaka Prefecture University, 1-1 Gakuen-cho, Nakaku, Sakai, Osaka 599-8531, Japan}
\affil{$^5$Solar-terrestrial Environment Laboratory, Nagoya University, Furo-cho, Chikusa-ku, Nagoya, Aichi 464-8601, Japan}
\affil{$^6$Department of Physics, Faculty of Science, Hokkaido University, N10W8, Kita-ku, Sapporo 060-0810, Japan}
\email{ohama@a.phys.nagoya-u.ac.jp}



\begin{abstract}
\citet{furukawa09} reported the existence of a large mass of molecular gas associated with the super star cluster Westerlund 2 and the surrounding HII region RCW49, based on a strong morphological correspondence between NANTEN2 $^{12}$CO(J=2-1) emission and Spitzer IRAC images of the HII region.  We here present temperature and density distributions in the associated molecular gas at $\sim$3.5 pc resolution, as derived from an LVG analysis of the $^{12}$CO(J=2-1), $^{12}$CO(J=1-0) and $^{13}$CO(J=2-1) transitions. The kinetic temperature is as high as $\sim$60-150 K within a projected distance of $\sim$5-10 pc from Westerlund 2 and decreases to as low as $\sim$10 K away from the cluster. The high temperature provides robust verification that the molecular gas is indeed physically associated with the HII region, supporting Furukawa et al.'s conclusion. The derived temperature is also roughly consistent with theoretical calculations of photo dissociation regions (PDRs), while the low spatial resolution of the present study does not warrant a more detailed comparison with PDR models. We suggest that the molecular clouds presented here will serve as an ideal laboratory to test theories on PDRs in future higher resolution studies.
\end{abstract}


\keywords{ISM: clouds - open clusters and associations: individual (Westerlund 2) - HII regions: individual (RCW 49)}



\section{Introduction}
The process of massive star formation is still poorly understood, partly because the majority of massive star forming regions are in highly confused and crowded environments. Study of the ISM around young stellar clusters can help to clarify the astrophysical activity in such regions, and can provide invaluable information on the earliest stages of massive star formation.

Super star clusters (SSCs) are massive clusters with stellar densities exceeding 10$^4$ stars pc$^{-3}$ in their cores \citep{johnson05}. SSCs strongly impact the Galactic ecosystem by supplying the ISM with energy and momentum via strong UV radiation, stellar winds and supernova explosions. SSCs are very rare in the Galaxy and there are only several known to date, although other galaxies harbor more \citep[e.g.][and references within]{kornei09}. Three are located in the Galactic center \citep[the Arches cluster, the Quintuplet cluster and the Central cluster;][]{figer99, genzel03} and the other two in the Galactic disk (Westerlund 1 and Westerlund 2; hereafter Wd2).

Wd2 is located at ($\textit{l}$, $\textit{b}$)=(284.2, -0.33), close to the tangent of the Carina Arm. Of the five known SSCs in the Galaxy, it is the only one for which associated molecular clouds have been identified (Furukawa et al. 2009, see also Dame et al. 2007). The estimated age of the cluster is 2-3 Myr (Piatti et al. 1998), and estimates of the total stellar mass range from $\sim$1$\times$10$^{4}$ to $\sim$3$\times$10$^{4}$$\textit{M}$$_{\odot}$ \citep{ascenso07, whitney04}, with several thousand $\textit{M}$$_{\odot}$ in the form of stars of 1$\leq$$\textit{M}$$_{\odot}$ $\leq$ 120 $\textit{M}$$_{\odot}$ \citep{ascenso07, rauw07}. This rich stellar population includes 12 O stars and 2 WR stars. 

Wd2 ionizes an extensive and luminous HII region, RCW49. This has recently been observed with the Infrared Array Camera (IRAC) on the Spitzer Space Telescope as a part of the GLIMPSE survey \citep{benjamin03}. \citet{churchwell04} present IRAC images of a dust emission and discuss polycyclic aromatic hydrocarbon (PAH) in the region, which is likely excited by the strong ultraviolet/optical radiation from the central cluster 
 Studies of IR point sources suggest that there are about 300 Young Stellar Objects (YSOs) more massive than 2.5 $\textit{M}_{\odot}$ in RCW49 \citep{whitney04}. 

In addition, there are observations in high energy bands in the region. RCW49 has been detected in X-rays in the 0.2 to 4.5 keV range by \citet{hertz84} and \citet{goldwurm87}, in the 0.1 to 2.4 keV range by \citet{belloni94}, and 468 X-ray point sources have been detected with Chandra by \citet{tsujimoto07}. An extended X ray feature of mainly thermal origin has also been detected toward Wd2 by the Suzaku X ray satellite \citep{2009arXiv0907.4750F}. Observations with the HESS TeV gamma-ray telescope reveal an extended TeV gamma-ray source, HESS J1023-575, toward RCW49 \citep{aharonian07}. Emission scenarios include the interaction of cosmic ray particles accelerated by stellar wind or supernova with molecular gas \citep{2009PASJ...61L..23F}.

Molecular observations provide an invaluable addition to this rich set of multi-wavelength data. \citet{dame07} used $^{12}$CO(J=1-0) survey data taken with the CfA 1.2m telescope at a resolution of 8.8$'$ to identify an extended giant molecular cloud (GMC) at a velocity of $\sim$11 km s$^{-1}$ as a possible parent GMC of Wd2. More recently, Furukawa et al. (2009; hereafter Paper I) carried out $^{12}$CO(J=2-1) observations at a resolution of $90\arcsec$ with the NANTEN2 4m telescope of Nagoya University. These authors present morphological evidence that the HII region is associated with two complexes of molecular gas, one in the velocity range $-$11 to 6 km s$^{-1}$ and the other between 11 and 21 km s$^{-1}$. These authors argued for the robust association of these clouds with Wd2, and concluded that a loose association with the more extended 11 km s$^{-1}$ cloud of Dame (2007), while possible, is not well supported by the data. The kinematic distance to these molecular clouds is estimated to be 5.4$^{+1.1}_{-1.4}$ kpc by using the rotation curve of \citet{brand93}, and the masses are estimated to be around 10$^5$ $\textit{M}$$_{\odot}$ each. Based on the mass and spatio-velocity structure of the gas, Paper I suggested that a collision between the two clouds may have triggered formation of the stellar cluster Wd2.

In this paper we present an analysis of molecular data of the $^{12}$CO(J=2-1), $^{12}$CO(J=1-0) and $^{13}$CO(J=2-1) emission lines. Section 2 details the observations themselves. Section 3 presents maps, derives line ratios and carries out a Large Velocity Gradient (LVG) analysis in order to constrain temperatures and densities in the clouds. We find strong evidence from the derived temperature distributions to support the conclusions of Paper I. In section 4 we compare our molecular gas temperatures with the results of simple PDR models, before summarizing our results in section 5.

\section{Observations}

This study makes use of new $^{13}$CO(J=2-1) observations, the recent $^{12}$CO(J=2-1) data from Paper I and archival $^{12}$CO(J=1-0) data from the NANTEN CO Galactic Plane Survey \citep{mizuno04}. Both J=2-1 transitions were observed with the NANTEN 2 4m sub-mm/mm telescope of Nagoya University, which is installed at Pampa La Bola (4800 m above sea level), in northern Chile.

	Observations of the $^{12}$CO(J=2-1) and $^{13}$CO(J=2-1) lines were conducted in February 2008 and December 2008, respectively. The backend was a 4 K cooled Nb SIS mixer receiver, and the single-side-band (SSB) system temperature was $\sim$200 K in $^{12}$CO(J=2-1) and $\sim$100 K in $^{13}$CO(J=2-1), including the atmosphere toward the zenith. We used two 2048-channel acousto-optical spectrometers (AOS), resulting in a bandwidth of 390 km s$^{-1}$ and a channel resolution of 0.19 km s$^{-1}$ at 230 GHz. The telescope half-power beam width was 90$\arcsec$ at 230 GHz. The pointing was checked regularly on Ori KL, $(\alpha, \delta)_{B1950}=(5^h~32^m~47.0^s, -5^{\circ}~24'~22\arcsec)$, and the applied corrections were always smaller than 20$\arcsec$; usually less than 10$\arcsec$. The target region was observed between elevations of 30 degrees and 60 degrees. All observations were conducted using an on-the-fly (OTF) mapping technique, and each spatial region was mapped several times in different scanning directions to reduce scanning effects. The final pixel size of the gridded data is 30$\arcsec$. The effective integration time per pixel is $\sim$2 s in $^{12}$CO(J=2-1) and $\sim$4 s in $^{13}$CO(J=2-1), resulting in rms noise levels per 0.19 km s$^{-1}$ channel of $\sim$1.5 K and $\sim$0.5 K, respectively. The standard  source $\rho$ Oph East, $(\alpha, \delta)_{B1950}=(16^h~29^m~20.9^s, -24^{\circ}~22'~30\arcsec)$, was observed for intensity calibration. Calibration uncertainties of 12\% in $^{12}$CO(J=2-1) and 8\% in $^{13}$CO(J=2-1) are estimated from the daily variation in its observed brightness temperature.


	The $^{12}$CO(J=1-0) line was observed by position switching as part of the NANTEN Galactic Plane Survey. The beam size at 115 GHz was 2.6$'$, and the data are gridded to a 2$'$ grid with a velocity resolution of 0.1 km s$^{-1}$. The rms noise per channel is $\sim$0.9 K, and calibration uncertainty is estimated at $\sim10$\%.
		
\section{Results}
\subsection{Distributions}

\begin{figure}
\epsscale{1}
\plotone{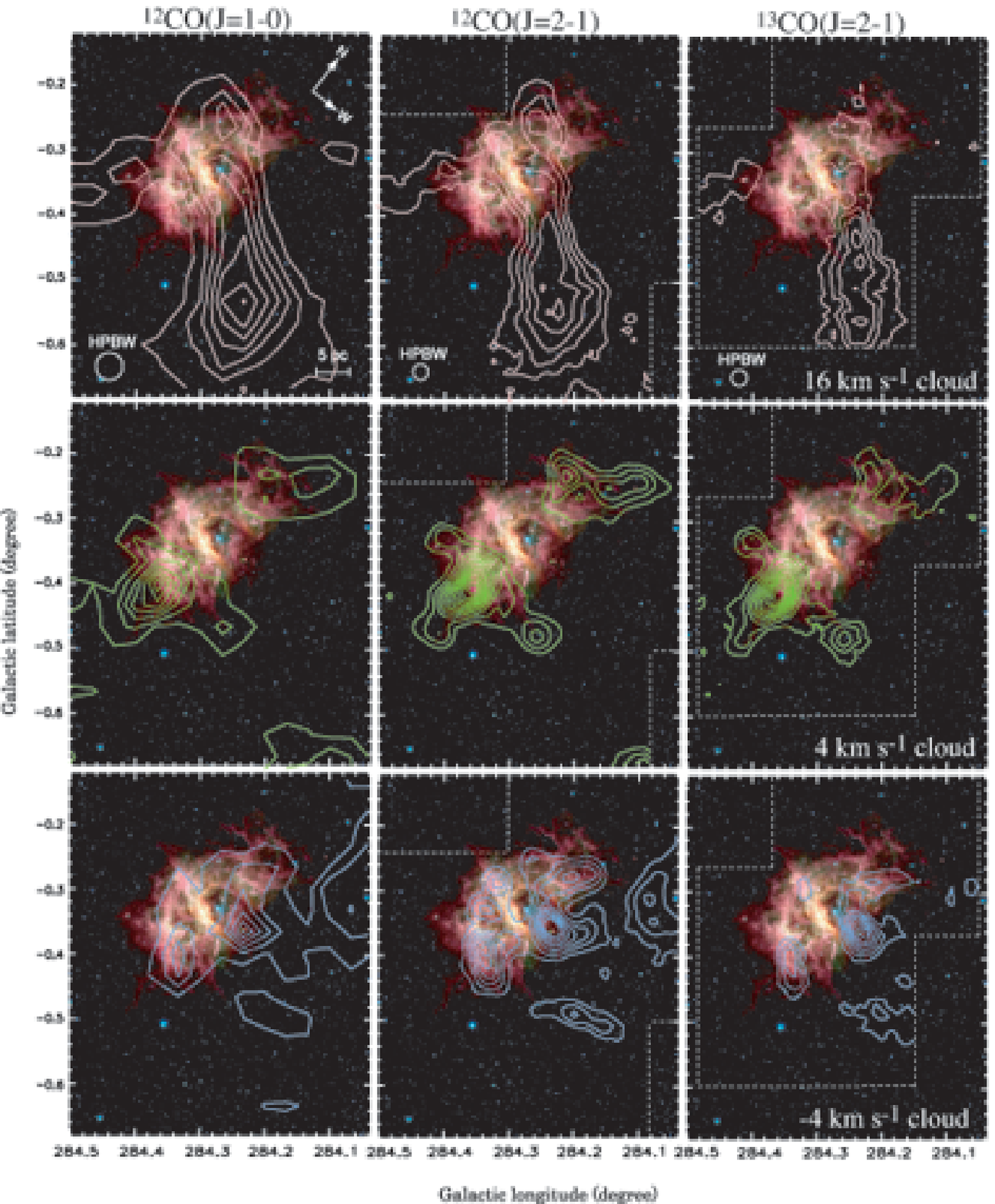}
\caption{CO integrated intensity contours overlaid on Spitzer three color image of RCW49. The left column shows  $^{12}$CO(J=1-0) data, the middle column $^{12}$CO(J=2-1) and the right column $^{13}$CO(J=2-1). The top, middle and bottom rows show emission in velcity ranges of 11.0 to 20.9 km s$^{-1}$, 1.2 to 8.7 km s$^{-1}$ and $-$11.0 to 0.3 km s$^{-1}$ respectively. Contour levels are 10 + 10 K km s$^{-1}$ for $^{12}$CO(J=1-0), 15 + 10 K km s$^{-1}$ for $^{12}$CO(J=2-1) and 2 + 2 K km s$^{-1}$ for $^{13}$CO(J=2-1). Here both $^{12}$CO(J=2-1) and $^{13}$CO(J=2-1) datasets have been convolved with a gaussian of FWHM 45".\label{fig1}}
\end{figure}

Figure \ref{fig1} shows the integrated intensity distributions of the three CO transitions, overlaid on a three-color IR image from the Spitzer/IRAC GLIMPSE survey. Following Paper I, we show the molecular gas distribution in velocity ranges of $-$11 to 0 km s$^{-1}$, 1 to 9 km s$^{-1}$ and 11 to 21 km s$^{-1}$ and hereafter refer to emission in these ranges as the $-$4 km s$^{-1}$, 4 km s$^{-1}$ and 16 km s$^{-1}$ clouds, respectively. However, it should be noted that the 4 km s$^{-1}$ and $-$4 km s$^{-1}$ clouds do not form completely distinct entities in the 3D data cubes, and that spectra in the $-$11 to 9 km s$^{-1}$ range are sometimes complex and blended.
	
As described in Paper I, the Northeastern half of the 16 km s$^{-1}$ cloud is located toward the central regions of RCW49. Notably, the Southeastern edge of the cloud, between $(l, b)=(284.33, -0.28)$ and $(284.27, -0.40)$, shows a strong correlation with a bright filamentary ridge of Spitzer emission $\sim2'$ Southeast of Wd2. The remainder of the cloud extends 0.2 degrees to the Southwest, extending fully outside the area of influence of the HII region. The 4 km s$^{-1}$ cloud shows an excellent morphological correlation with the large-scale shape of the HII region. Emission at this velocity range is entirely absent within a radius of 3$'$ of Wd2, and at larger distances from the cluster the molecular gas fans out to the North and South, matching well with the IR-bright gas. The $-$4 km s$^{-1}$ cloud distribution also matches well with the IR nebula and in particular we note that a prominent CO clump centered on (284.22, $-$0.29) appears to be associated with a number of IR point sources, one of which has been identified as a YSO of spectral type B2 \citep{churchwell04}.
	
	New to the present study is the addition of $^{13}$CO(J=2-1) data from NANTEN2, as well as archival $^{12}$CO(J=1-0) data from the older NANTEN Galactic Plane Survey. All three lines follow very similar distributions at the present spatial resolution, with $^{13}$CO(J=2-1) emission detected towards the brighter regions of the $^{12}$CO clouds. This can be seen in Figure \ref{fig1}.
	
\subsection{Line Intensity Ratios}

	We convolve the $^{12}$CO(J=2-1) and $^{13}$CO(J=2-1) data with a Gaussian of FWHM 2.6$'$ in order to smooth the cubes to the resolution of the $^{12}$CO(J=1-0) beam. We then derive peak main beam temperatures, $\textit{T}$$_{mb}$ (K), and linewidths for the relevant emission components at each $^{12}$CO(J=1-0) pixel position by fitting either single or double Gaussian functions to the line profiles. Two ratios, $R_{2-1/1-0} = T_{mb}[^{12}$CO(J=2-1)$]/T_{mb}[^{12}$CO(J=1-0)$]$ and $R_{12/13} = T_{mb}[^{12}$CO(J=2-1)$]/T_{mb}[^{13}$CO(J=2-1)$]$, are then calculated. The uncertainties on these ratios are around 15\%, as calculated from the propagation of the intensity calibration errors, under the assumption that these errors are uncorrelated.

\begin{figure}
\epsscale{1}
\plotone{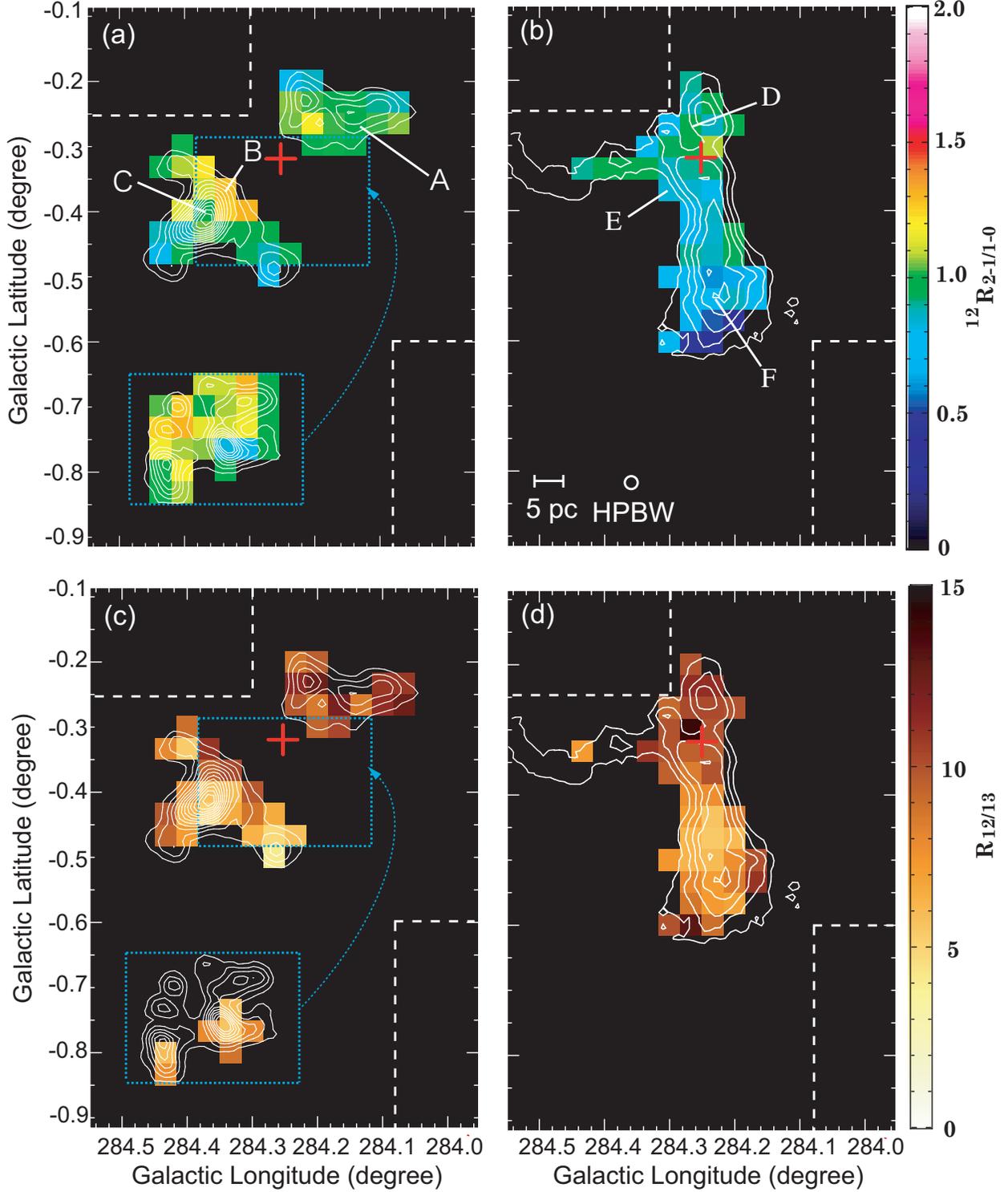}
\caption{Maps of the intensity ratios $R_{2-1/1-0}$ (top) and $R_{12/13}$ (bottom) overlaid with $^{12}$CO(J=2-1) integrated intensity contours from Figure \ref{fig1}. Panels (a) and (c) show the 4 km s$^{-1}$ cloud and panels (b) and (d) the 16 km s$^{-1}$ cloud. The -4 km s$^{-1}$ component is shown in the inset blue dotted box. The red cross marks the position of Wd2. Letters refers to the points displayed in Figure \ref{fig3}.\label{fig2}}
\end{figure}

Of these ratios the former is of most immediate interest, being a measure of the level of excitation of the molecular gas. We find that in the 4 km s$^{-1}$ cloud $R_{2-1/1-0}$ ranges from 0.42 to 1.34 while the 16 km s$^{-1}$ cloud shows values of between 0.36 and 1.09. In the $-$4 km s$^{-1}$ cloud, for which all three lines were significantly detected at only five 2$'$ pixels, 
ratios of between 0.80 and 1.34 are observed. The spatial distribution of $R_{2-1/1-0}$ is shown in panels (a) and (b) of Figure \ref{fig2}. It is striking that molecular gas at smaller projected distances from Wd2 generally shows higher ratios. In both the $-$4 and 4 kms$^{-1}$ clouds, molecular gas spatially coincident with the HII region shows markedly enhanced ratios, and in particular the southern part of the 4 km s$^{-1}$ cloud shows the highest ratios toward the cluster in the north but lower ones in the south.
This suggests that gas closer to the cluster is more highly excited, and is quite consistent with the association suggested by \citet{furukawa09}. 


Panels (c) and (d) of Figure \ref{fig2} show the spatial distribution of $R_{12/13}$. This ratio ranges from 4.1 to 13.0 across the entire molecular gas complex. Smaller values indicate higher optical depths and, as expected, are generally found towards CO emission peaks where column densities are larger.

\subsection{LVG analysis}

We may explore the issues of temperature and density more robustly by using a large velocity gradient (LVG) analysis \citep{goldreich74}. LVG analysis is a method of calculating a simplified radiative transfer code based on the approximate escape probability formalism of \citet{castor70}. We assume a spherically symmetric cloud of constant density and temperature with radial velocity distribution proportional to the radial distance from the cloud center. We then utilize the line ratios measured above to solve the equations of statistical equilibrium for the fractional population of the lowest 40 rotational levels of the ground vibrational level of CO. The velocity gradient, d$\textit{v}$/d$\textit{r}$, is taken to be 2.6 km s$^{-1}$ pc$^{-1}$, as estimated from the typical linewidth in a single 3.5 pc cell. In order to evaluate the dependence of temperature on the adopted parameters we tested three values of $dv/dr$: 1.2, 2.6 and 5.0 km s$^{-1}$ pc$^{-1}$, while keeping the other parameters fixed. We find that $T_k$ varies by about a factor of 1.6 for a factor of 2 change in d$\textit{v}$/d$\textit{r}$, indicating that the results do not depend strongly on d$\textit{v}$/d$\textit{r}$. The CO fractional abundance $\textit{X}$(CO) is taken to be 5 $\times$ 10$^{-5}$ \citep{sakamotophd}, and the $^{12}$CO/$^{13}$CO abundance ratio is taken to be 75 \citep{guesten04}. 
	
\begin{figure}
\epsscale{1}
\plotone{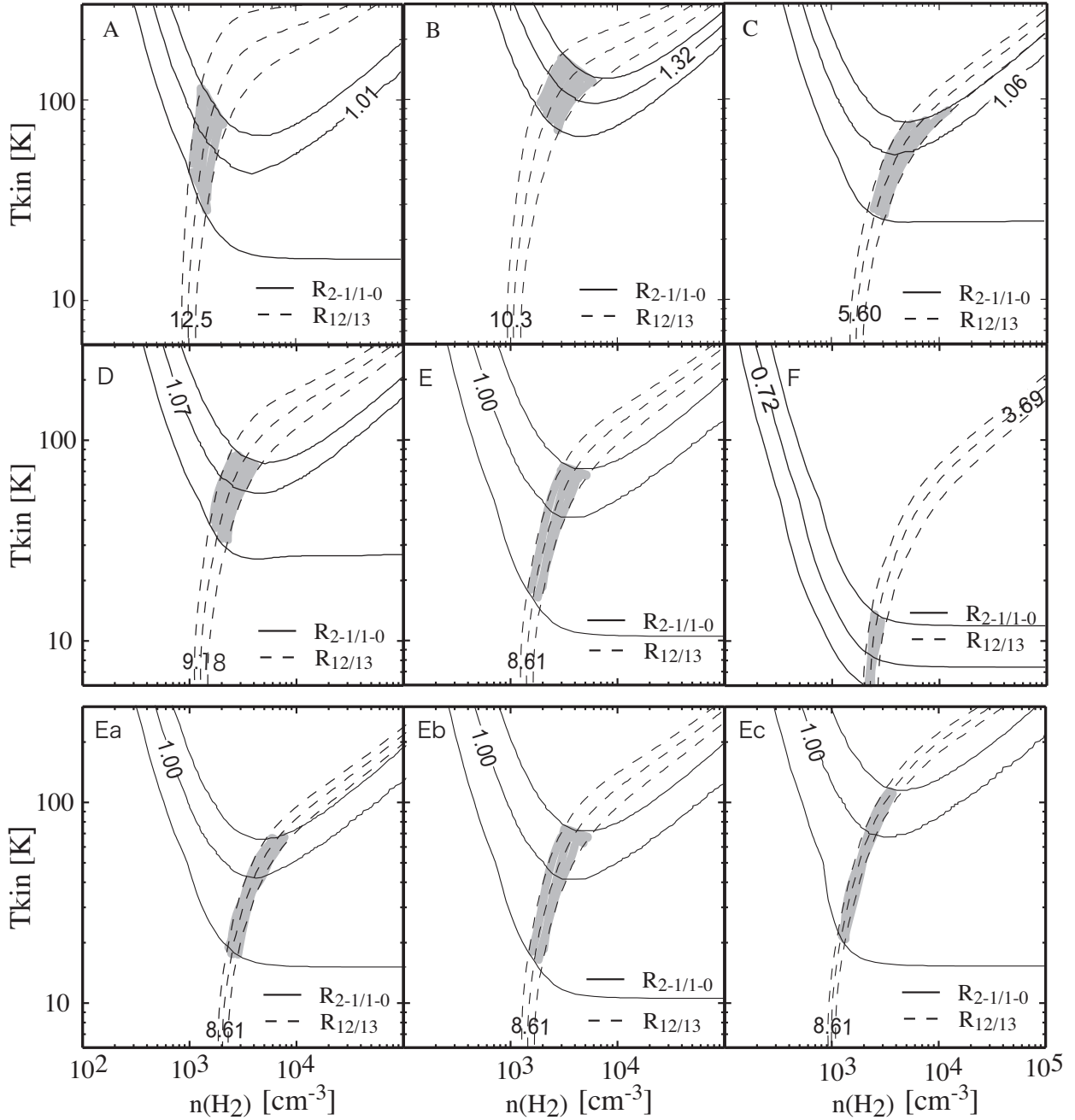}
\caption{Contour plots of the LVG solutions at points A to F, as marked in Figure \ref{fig2}. Solid and dashed lines mark solutions derived from the ratios $R_{2-1/1-0}$ and $R_{12/13}$ respectively. Here the outer lines indicate 15$\%$ calibration uncertainties, and shaded grey areas are where the two solutions coincide. Panels A-F assume  $\textit{X}$(CO) of 5$\times$10$^{-5}$ and d$\textit{v}$/d$\textit{r}$ of 2.6 km s$^{-1}$ pc$^{-1}$. Panels Ea, Eb and Ec show the solutions for point E under the assumption of values of X/(d$\textit{v}$/d$\textit{r}$) of 1.0$\times$10$^{-5}$, 2.6$\times$10$^{-5}$ and 4.0$\times$10$^{-5}$ (km s$^{-1}$)$^{-1}$ pc, respectively.\label{fig3}}
\end{figure}

\begin{figure}
\epsscale{1}
\plotone{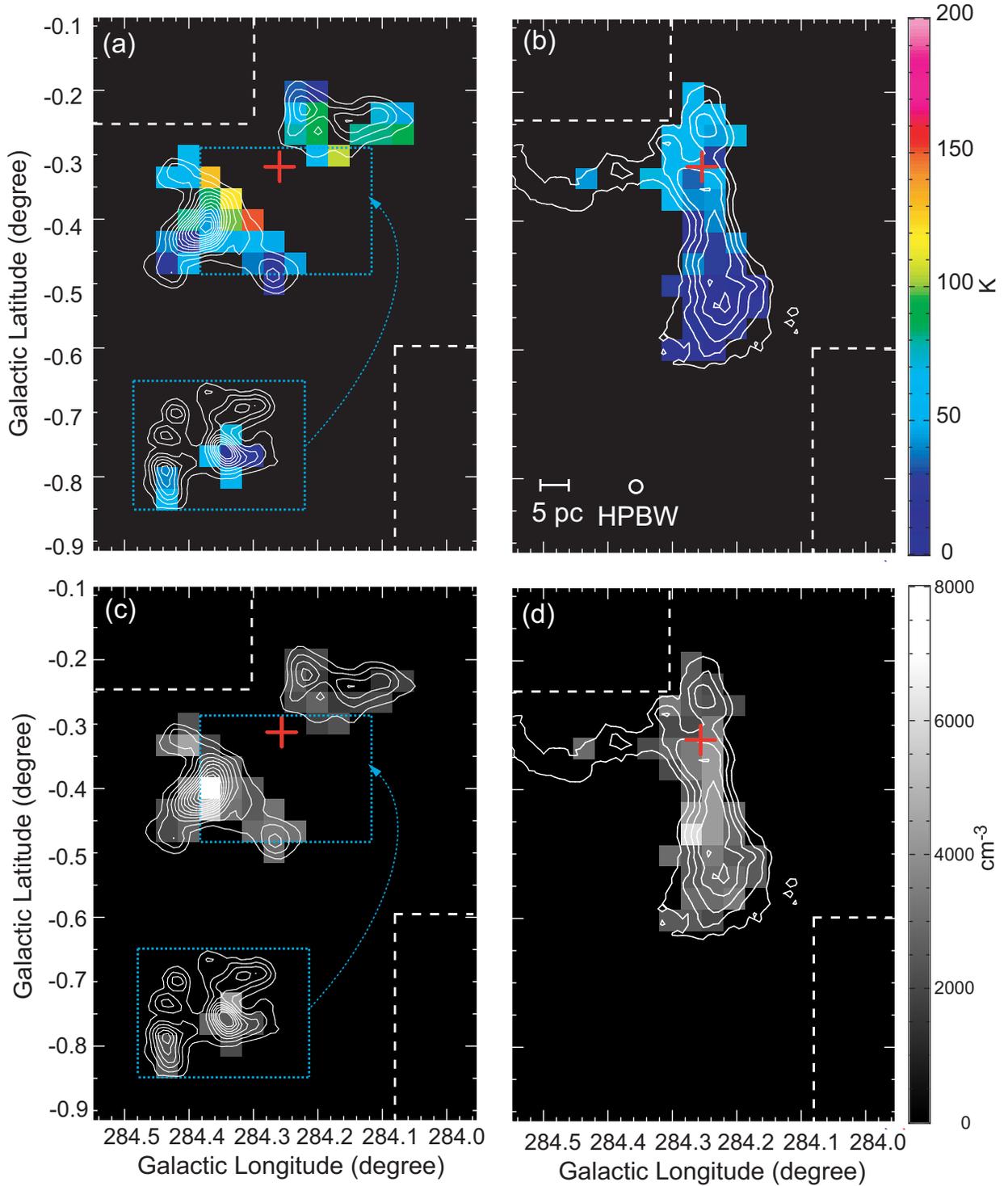}
\caption{LVG-derived temperature (panels (a) and (b)) and density (panels (c) and (d)) distributions overlaid with $^{12}$CO(J=2-1) integrated intensity contours from Figure \ref{fig1}. Panels (a) and (c) show the 4 km s$^{-1}$ cloud and panels (b) and (d) the 16 km s$^{-1}$ cloud. The -4 km s$^{-1}$ component is shown in the inset dotted box. The red cross marks the position of Wd2.\label{fig4}}
\end{figure}

\begin{figure}
\epsscale{1}
\plotone{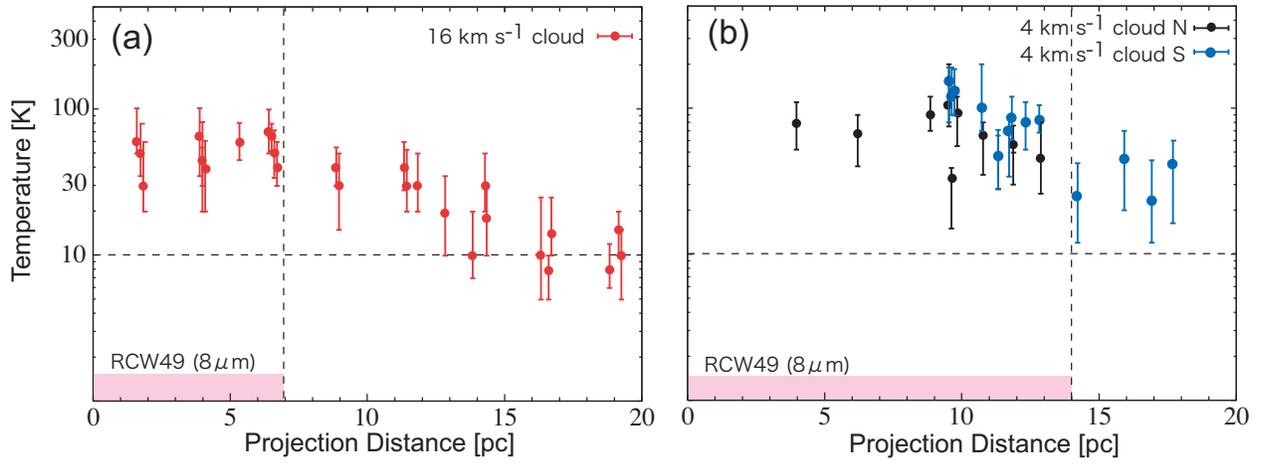}
\caption{Plot of relation between LVG-derived kinetic temperature and projected distance from Wd2. Panel (a): red points show the 16 km s$^{-1}$ cloud. Panel (b): Black and blue points show the Northern part of the 4km s$^{-1}$ cloud and  the Southern part of the 4 km s$^{-1}$ cloud, respectively. Pink bars show the extent of the IR nebula RCW49.\label{fig5}}
\end{figure}

Figure \ref{fig3} shows solutions for kinetic temperature, $\textit{T}$$_{k}$, and number density, $\textit{n}$$_{0}$, at selected positions, labelled A to F in Figure \ref{fig2}. In Figure \ref{fig4} we plot the spatial distributions of $T_k$ and $\textit{n}$$_0$ over the entire extent of the clouds, only excluding points for which no $^{13}$CO(J=2-1) emission was detected. (In the case of the -4 km s$^{-1}$ cloud $^{13}$CO(J=2-1) is only significantly detected at seven pixels and we therefore exlude it from the following discussion.) It may be seen that for both the 16 km s$^{-1}$ and 4 km s$^{-1}$ clouds the kinetic temperature is enhanced toward the Spitzer IR nebula and at smaller projected distances from the cluster. This is especially true in the case of the 4 km s$^{-1}$ cloud, which shows temperatures as high as 70$-$150 K on the side facing Wd2. $\textit{T}$$_{k}$ in the 16 km s$^{-1}$ cloud is generally lower, reaching peak values of 30$-$70 K toward the IR nebula. However, these values are still significantly enhanced with respect to the typical temperatures of $\sim$10 K observed in molecular clouds without a local heat source. Moreover, the 16 km s$^{-1}$  cloud shows a clear decrease in $\textit{T}$$_{k}$ with projected distance from Wd2, decreasing to to 10 K beyond the IR nebula at distances of 15$-$20 pc from the central cluster. The temperature in the Southern part of 4 km  s$^{-1}$ cloud also drops slightly to 30$-$50 K beyond the IR nebula at projected distances of 10$-$20 pc from the cluster, while the Northern part of 4 km  s$^{-1}$ cloud located toward the IR nebula within 10 pc of Wd2 emains high in temperature. This is illustrated in Figure \ref{fig5}, which shows plots of $\textit{T}$$_{k}$ as a function of projected distance from Wd2. This relationship supports the association between the molecular clouds and the stellar cluster suggested in Paper I, while the lower $\textit{T}$$_{k}$ in the 16 km s$^{-1}$ cloud compared to the 4 km s$^{-1}$ cloud may suggest that it is located further from the cluster. We also note that the 4 km s$^{-1}$ cloud follows more closely the shape of the IR nebula as it extends in the North-South direction, whereas the 16 km s$^{-1}$ cloud is elongated to the Southwest, with only its Northeastern most portions coincident with the Spitzer emission. The extent of the IR emission in each cloud with increasing distance from Wd2 is indicated in Figure 5 and explains why the high temperature regions in the 4 km s$^{-1}$ cloud extend to larger radii than in the 16 km s$^{-1}$ cloud.

In general it is sensible to be cautious about the accuracy of the quantity $\textit{T}$$_{k}$ when the energy levels employed are not as high as the temperatures derived. In the present case the J=1 and 2 levels of CO are at 5.5 K and 16 K above ground level, respectively. Nevertheless, the temperature ranges indicated in Figure 3 demonstrate that the analysis is accurate enough to conclude that the kinetic temperatures are significantly higher than 10 K, the typical temperature of molecular gas without an extra heat source other than the general interstellar radiation field and cosmic ray sea.

Number densities are typically $\sim$3000 cm$^{-3}$ in both clouds, with the Southern part of of the 4 km s$^{-1}$ cloud showing higher densities than the Northern part. In particular, the $^{12}$CO(J=2-1) integrated intensity peak has an estimated $\textit{n}$$_0$ of $\sim$8000 cm$^{-3}$. It is also notable that the 16 km s$^{-1}$ cloud shows a slight density depression close to Wd2, suggesting gas dispersal by the stellar cluster.

In PDR regions, the abundance of $^{13}$CO may change by a factor up to 2.0 relative to  $^{12}$CO in regions of $A_V$ less than $\sim$1 magnitude, where chemistry becomes complicated by partial ionization \citep{2009A&A...503..323V}. The present clouds have significantly higher $A_V$, of up to 10 mag, and the fraction of the positions with $A_V$ less than 1.5 mag is only $\sim$ 15\%, suggesting that the effects of $^{13}$CO abundance variation may not be important. In order to test the effects of the $^{12}$CO/ $^{13}$CO ratio  we examine cases in which this ratio varies between 40 and 150, instead of the 75 assumed in the original analysis. We find that the derived temperature is still significantly high; for example, $\textit{T}$$_{k}$ = 70 K at (Position A) in the -4 km s$^{-1}$ cloud falls in the range 45-90 K, significantly higher than 10 K. Considering the large values of $A_V$ noted above, this temperature range gives conservative limits. Thus we argue that the effect of the PDR on $^{13}$CO abundances does not alter the key result that temperatures towards the HII region are high, and the association of the clouds remains valid.

\section{Discussion}
\subsection{Comparison with PDR Models}

RCW49 is a classic example of a photodissociation region (PDR), in which the dense molecular ISM and dust are being heated by a strong FUV field. It is therefore instructive to make a simple comparison of the PDR surface temperatures and the molecular gas temperatures obtained above.

We make use of the standard model of \citet{kaufman99}, which parameterizes the PDR surface temperature of a constant density cloud in terms of the number density of hydrogen nuclei, $\textit{n}$$_0$, and the FUV (6 eV $\le$ E $\le$ 13.6 eV) field at the cloud surface, $\textit{G}$$_0$. We take $\textit{n}$$_0$ as $\sim$6$\times$10$^{3}$ cm$^{-3}$, based on the LVG analysis above, and estimate $\textit{G}$$_0$ from the Far Infra Red (FIR) luminosity of the nebula. Here we utilize high-resolution IRAS 60 $\mu$m and 100 $\mu$m data from the Infrared Processing and Analysis Center (IPAC). The filter properties of IRAS allow us to combine these data sets to obtain the FIR intensity, $\textit{I}$$_{FIR}$, between 42.5 $\mu$m and 122.5 $\mu$m, via the following relation (Helou et al. 1988; Nakagawa et al. 1998): 
\begin{equation}
	I_{FIR} = 3.25 \times 10^{-5}\times F(60\mu m)+1.26 \times 10^{-5} \times F(100\mu m)
\end{equation}
Where $\textit{F}$(60 $\mu$m) and $\textit{F}$(100 $\mu$m) are the 60 $\mu$m and 100 $\mu$m fluxes in MJy sr$^{-1}$.
Assuming that the FUV energy absorbed by the grains is reradiated in the FIR, we then estimate the FUV flux G$_{0}$ impinging onto the cloud surface from:
\begin{equation}
G_0 = 4 \pi \times I_{FIR} /1.6 \times 10^{-3}
\end{equation}
Where $\textit{G}$$_0$ is in units of the Habing Field, 1.6$\times$10$^{-3}$ ergs cm$^{-2}$ s$^{-1}$. Following \citet{kramer08}, this formula assumes that photons with E $\le$ 6 eV contribute about half of the total heating \citep{tielens85}, but also that the total re-radiated bolometric IR luminosity is a factor of two larger than $\textit{I}$$_{FIR}$ \citep{dale01}. These factors approximately cancel. 

	At the hottest location in the 4 km s$^{-1}$ molecular cloud $(l, b)=(284.33, -0.366)$, $\textit{F}$(60 $\mu$m) and $\textit{F}$(100 $\mu$m) are 2.3$\times$10$^{4}$ and 1.7$\times$10$^{4}$ MJy sr$^{-1}$, respectively. $\textit{G}$$_0$ is therefore estimated to be 7.5$\times$10$^{3}$ ergs cm$^{-2}$ s$^{-1}$ sr$^{-1}$. Comparing with Figure \ref{fig1} of \citet{kaufman99}, the PDR surface temperature is thus estimated to be several hundred K. 

	This result, although slightly higher than the derived LVG temperatures in the molecular gas, is nonetheless highly consistent when we consider the fact that the observed CO emission will not originate from the active cloud surface. Instead it is likely to originate from deeper, and therefore presumably colder, regions of the cloud, which have not yet been penetrated by the dissociating radiation.

	The above comparison is necessarily crude. Future high-resolution studies with instruments such as ALMA may resolve molecular gas emission down to the size scales as the structures observed by GLIMPSE, and pave the way to more detailed model comparisons. 
	
\subsection{Embedded YSOs in the -4km s$^{-1}$ clump}

	Finally, we shall also briefly mention the prominent clump in the -4 km s$^{-1}$ cloud located at $(l,b)=(284.23, -0.36)$. This clump shows hints of higher values of $R_{2-1/1-0}$ and $\textit{T}$$_{k}$ in its periphery than in its center (see Figure 2a and 4a), suggesting that its outer regions are heated by the UV photons of Wd2. 
The clump itself also appears to be associated with several internal YSOs, with a maximum spectral type of around B2 \citep{2005RMxAC..23...53C, whitney04}. The total luminosity of a single ZAMS B2 star may be estimated from the evolutionary tracks of \citet{iben65} to be around 2.1$\times$10$^{37}$ erg s$^{-1}$. For comparison, the cooling rate of the molecular clump is estimated to be 1$\times$10$^{34\pm0.7}$ ergs s$^{-1}$, where we follow the cooling rate equation of \citet{1978ApJ...222..881G} and assume a kinetic temperature of 20 K, a density of 4$\times$10$^{3}$ cm$^{-3}$ and approximate the clump as a sphere of radius 2.5 pc. This figure is of the order of 1\% of the estimated B2 star luminosity. Considering that the molecular feature is not bright in the Spitzer PAH bands, we suggest that the interior of the clump is unlikely to be primarily heated by the UV radiation from the central cluster, and that the embedded YSOs are more than sufficient to heat the gas to it's present temperature.  It is both desirable and important to obtain future detailed molecular observations at higher resolution to explore its detailed properties in order to better understand the star formation occuring within it.
	

\section{Summary}

We have carried out an analysis of three CO transitions -- $^{12}$CO(J=1-0),  $^{12}$CO(J=2-1) and $^{13}$CO(J=2-1) -- toward Westerlund 2, and derived temperature and density distributions for the two associated molecular cloud complexes identified in Paper I. The main conclusions are listed below. These conclusions reinforce Furukawa et al.'s suggestion that the molecular gas is physically associated with Wd2 and RCW49, and therefore provides support for their idea that a cloud-cloud collision may have triggered the formation of this remarkable cluster.

1. The ratio of peak main beam tempertures in the $^{12}$CO(J=2-1) and $^{12}$CO(J=1-0) lines, $R_{2-1/1-0}$, ranges from 0.42 to 1.34, and the ratio in the $^{13}$CO(J=2-1) and $^{12}$CO(J=1-0) lines, $R_{12/13}$, ranges from 4.1 to 11.8. In both the 4 km s$^{-1}$ and 16 km s$^{-1}$ clouds, regions toward the IR nebula and  closer to the cluster show enhanced values of $R_{2-1/1-0}$, indicating a higher level of excitement closer to the stellar heating source. $R_{12/13}$ ranges from 4.1 to 13.0 across the entire complex of associated gas.

2. We have carried out a Large Velocity Gradient (LGV) analysis to estimate temperatures and densities in the clouds. Kinetic temperatures derived are as high as from  $\sim$30 to $\sim$150 K toward the IR nebula within 10$-$15 pc of the cluster both in the 4 km s$^{-1}$ cloud and in the 16 km s$^{-1}$ cloud. These high temperatures require a local heat source, and provide robust verification of the physical association between the clouds and the HII region and stellar cluster, as suggested in Paper I.  Densities are typically around 3000 cm$^{-3}$, with peak values of around 8000 cm$^{-3}$.

3. The 16 km s$^{-1}$ cloud, which extends beyond the IR nebula as far as 20 pc from Wd2, shows a clear decrease in $\textit{T}$$_k$ with projected distance from the cluster. This, in addition to the high temperature toward the IR nebula, supports the association between it and the molecular cloud. The Southern part of the 4 km s$^{-1}$ cloud also shows a similar temperature decrease with distance, with lower temperatures observed outside the IR nebula at a projected distance of 15$-$20 pc from the cluster. Similarly, the Northern part of the 4 km s$^{-1}$ cloud, which is located toward the IR nebula and lies within 10-14 pc of Wd2, also shows generally high temperatures. The fact that the 16 km s-1 cloud shows lower temperatures than the 4 km s$^{-1}$ cloud suggests that it may be located at a somewhat larger distance from the cluster.

4. We have made use of simple PDR model calculations to estimate a PDR surface temperature of several hundred K. This result is considered to be consistent with the LVG-based temperature estimates from the CO emission, which should originate from deeper and colder regions of the cloud.

5. The embedded YSOs at ($\textit{l}$,$\textit{b}$)=(284.23, -0.36) provide more than sufficient energy to heat the -4km s$^{-1}$ molecular clump in which they are located.




\acknowledgments
\section{Acknowledgments}

NANTEN2 is an international collaboration of ten universities, Nagoya University, Osaka Prefecture University, University of Cologne, University of Bonn, Seoul National University, University of Chile, University of New South Wales, Macquarie University, University of Sydney and Zurich Technical University. The work is financially supported by a Grant-in-Aid for Scientific Research (KAKENHI, No. 15071203, No. 21253003, and No. 20244014) from MEXT (the Ministry of Education, Culture, Sports, Science and Technology of Japan) and JSPS (Japan Soxiety for the Promotion of Science) as well as JSPS core-to-core program (No. 17004). We also acknowledge the support of the Mitsubishi Foundation and the Sumitomo Foundation. This research was supported by the Grant-in-Aid for Nagoya University Global COE Program, ''Quest for Fundamental Principles in the Universe: from Particles to the Solar System and the Cosmos'', from MEXT. Also, the work makes use of archive data acquired with Spitzer Space Telescope and IRAS data gained with Infrared Processing and Analysis Center (IPAC). Spitzer is controlled by the Jet Propulsion Laboratory, California Institute of Technology under a contract with NASA.

\bibliographystyle{apj}
\bibliography{ohama09bib}
\clearpage

\clearpage

\clearpage

\clearpage

\clearpage

\end{document}